\documentstyle[aaspp, epsf]{article}

\lefthead{Beckman {\it et al. }}
\righthead{Arm and Interarm Scale Lengths}
\slugcomment{\today}

\def\deg{\ifmmode^\circ\else$^\circ$\fi}

\def\etal{{\it et al.~}}
\def\deg{$^{\rm o~}$}
\def\bck{\hskip-0. 35em}
\def\min#1{\ifmmode  {^{\prime}}                           
            \else    {$^{\prime}$}\fi
            \ifcat, #1{\bck}\else\null\fi\ #1}
\def\deg{\ifmmode {^{\rm o}}              
         \else {$^{\rm o}$}\fi}
\def\sec{\ifmmode {^{\prime\prime}~}       
         \else {$^{\prime\prime}~$}\fi}

\begin{document}
\title{Scale Lengths in Disk Surface Brightness as probes of Dust 
Extinction in 3 Spiral Galaxies:  M~51,  NGC~3631 and M~100 }

\author{J. E.  Beckman}
\affil{Instituto de Astrof\'\i sica de Canarias}
\authoraddr{E-38200 La Laguna, 
  Tenerife, Spain}

\author{R. F.  Peletier}
\affil{Kapteyn Astronomical Institute}
\authoraddr{Postbus 800,  9700~AV~~Groningen,  Netherlands}

\and

\affil{Instituto de Astrof\'\i sica de Canarias}
\authoraddr{E-38200 La Laguna, 
  Tenerife, Spain}

\author{J. H.  Knapen}
\affil{D\'epartement de Physique,  Universit\'e de Montr\'eal}
\authoraddr{C. P. 6128, 
  Succursale Centre Ville, Montreal, Quebec, H3C 3J7, Canada; and
Observatoire de Mont M\'egantic}

\and

\affil{Instituto de Astrof\'\i sica de Canarias}
\authoraddr{E-38200 La Laguna, 
  Tenerife, Spain}

\author{R. L. M. Corradi}
\affil{Instituto de Astrof\'\i sica de Canarias}
\authoraddr{E-38200 La Laguna, 
  Tenerife, Spain}

\author{L. J. Gentet}
\affil{Department of Physics,  Imperial College}
\authoraddr{London, SW7 2BZ, 
  U. K. }

\and

\affil{Instituto de Astrof\'\i sica de Canarias}
\authoraddr{E-38200 La Laguna, 
  Tenerife, Spain}

\begin{abstract}

We have measured the radial brightness distributions in the disks of
three nearby face-on 
spirals: M~51, NGC~3631, and NGC~4321 (M~100) in the
photometric bands $B$ through $I$, with the addition of the $K$ band
for M~51 only.  The measurements were made by averaging azimuthally, in
three modes, the two-dimensional surface brightness over the disks in
photometric images of the objects in each band: (a) over each disk as
a whole, (b) over the spiral arms alone, and (c) over the interarm
zones alone.  From these profiles scale-lengths were derived for
comparison with schematic exponential disk models incorporating
interstellar dust. 
These models include both absorption and
scattering in their treatment of radiative transfer. 
The model fits show 
that the arms exhibit greater
optical depth in dust than the interarm zones.  The average fraction
of emitted stellar light in $V$ which is extinguished by dust within 3
scale-lengths of the center of each galaxy does not rise above 20\% in
any of them.  We show that this conclusion is also valid for models 
with similar overall quantities of dust, but where this is concentrated
in lanes. These can also account for the observed scale-lengths,
and their variations.

\end{abstract}


\section{Introduction}

Photometric mapping of spiral galaxies yields both detailed morphology
and a set of global physical parameters.  Azimuthally averaged
profiles of the surface brightness can usually be decomposed into a
central bulge and an exponential disk, whose relative importance is a
function of morphological type.  In this paper we will concentrate  
on the properties of the disks which will be characterized
by their observed exponential scale-lengths.  From the fairly
extensive detailed studies which have been published to date some
general conclusions have been drawn which are still subject to
empirical and theoretical debate. Most spiral galaxies have disks with
a broadly exponentially declining radial surface brightness
distribution (e.g. Kent 1984, de Jong 1995). 
It seems to be emerging that the scale-lengths observed
tend to decrease systematically, but not strongly, with increasing
wavelength, an effect which can be attributed to extinction by
moderate amounts of dust, with radial metallicity gradients as a
possible contributing factor (Elmegreen \& Elmegreen 1984; Prieto \etal
 1992; Evans 1994; Peletier \etal 1994, Beckman \etal 1995, de Jong 1995)
The disk surface brightness extrapolated to the axis of a
galaxy, $\mu_0$, lies within a narrow range, for a given photometric
band, which for the $B$ band is $\mu_0$ = 21.6 ($\pm$ 0.3)
mag~arcsec$^{-2}$ (Freeman 1970). For bands further into the red the
range is larger, and the central disk surface brightness depends more
strongly on spectral type (Peletier \etal 1994).  These studies of
disks have used one-dimensional, azimuthally averaged profiles to
derive the parameters cited above.  Galaxy disks are not, however,
axisymmetrical; they almost invariably show structures, such as arms
and bars. It is obvious from inspection of galaxy images that the arm
surface brightnesses are higher at all optical and NIR wavelengths
than those of the inter-arm zones of the disk, that arms often have
dust lanes at their edges, and that the arms are sites of more
concentrated star formation.  Because of the latter two conditions,
the brightness profile of the interarm zones may well be a more
faithful tracer of the underlying stellar mass distribution, and thus
offer a more appropriate way to determine the properties of the 
exponential disk. In an azimuthally averaged profile of a complete
galaxy disk the arm contribution may well be dominant, especially in
the shorter wavelengths, $B$ and $V$ bands.  Dust will affect the arms
and the interarm disk to different degrees at different wavelengths,
and with different geometries.  For these reasons we considered it a
valuable exercise to separate the arms from the inter-arm zones, and
to perform separate profile analyses. 
In a related paper (Knapen \& Beckman 1996), radial profiles in several
gas and star formation tracers were studied in detail for NGC~4321, for
arm and interarm regions separately, with special emphasis on the role
of the atomic hydrogen in the massive star formation process. In the
present paper we concentrate on dust, and present our results for three
well-studied nearby late-type spirals: M~51
(NGC~5495), NGC~3631, and NGC~4321 (M~100).  In Section 2 we describe
the data sets we used, and the methods of analysis applied, in Section
3 we present the observational results, which we compare, in Section
4, to the predictions of a set of theoretical models which incorporate
absorption and scattering in their extinction parameters.  In Section
5 we draw conclusions, and deal semi-quantitatively with the
consequences of departures from homogeneously exponential radial dust
distributions.

\section{The data and its treatment}

The observations were obtained from three different sources: two of
the galaxies were observed specifically for the purpose of the present
study, while for the third we used an existing set of images.  For M~51
we used images in the $B$, $R$, and $I$ optical bands kindly supplied
to us by H.-W. Rix. The acquisition and reduction of these images is
described by Rix and Rieke (1993).  They cover the inner disk of the
galaxy out to a brightness level of $\mu_B$ = 23 mag~arcsec$^{-2}$,
in
the $B$ band, with a scale of 1.78 arcsec~pixel$^{-1}$.
The images in $B$, $V$, $R$ and $I$ of NGC~3631 were obtained
in service time at the Cassegrain focus of the 1m Jacobus Kapteyn
Telescope (JKT) on La Palma, in January 1994, by P. Rudd and M. Asif.
They were taken with a 1280 $\times$ 1180 pixel EEV CCD camera, with
pixels subtending 0.31\sec\ , which implies a field of view of some
6\min $\times$ 6\min\ . This is satisfactory to cover the galaxy down
to $\mu_B$ = 24 mag~arcsec$^{-2}$,
as NGC~3631 is not especially large in angle.
We obtained $B$, $V$, $R$ and $I$ images
of NGC~4321, also in service time, at the prime focus of the 2.5m
Isaac Newton telescope (INT) on La Palma.  The basic reduction of the
images was described in Knapen \etal  (1993). 
Photometric calibration was performed via aperture 
photometry from the literature. For NGC~3631 we only found aperture photometry
in the $V$-band. The other colors were calibrated assuming at 
r=100\sec\ :
$B-V$ = 0.70, $V-R$ = 0.45 and $R-I$ = 0.50. NGC~4321 was calibrated
using aperture photometry in $B$, and assuming at r=100\sec\ :
$B-V$ = 0.90, $V-R$ = 0.55 and $R-I$ = 0.65. 
In Plate 1 we present real color
images of the three galaxies, produced by combining the images
in $B$, $V$ and $I$ ($B$, $R$ and $K$ for M~51).
In order to extract the radial profiles we first removed
bright foreground stars from our images, defining the areas affected
by the stars, and replacing the pixel values by blanks. The next step
was to produce, for each galaxy and in each band, two separate images,
one consisting only of the arm regions and the other of the inter-arm
regions.  To do this we first made a fit to the radial profile of the
I-band image by azimuthally averaging the light in elliptical annuli,
whose axes and ellipticities followed those of the disks as a
whole. We then defined the arms as the regions in which the intensity
was more than 7\% larger that the fitted profile value.  The
complexity of the inner parts of the galaxies ( out to some 20-30\sec\
from their nuclei) led us to define the arms here in the same way.
Also at large radii the definition of the arms was slightly revised 
interactively.
Since placing the boundary between arm and inter-arm regions is
somewhat subjective at least two authors checked the procedure
independently, with very good agreement in each case.  We used the $I$
band to produce a mask separating the arms from the inter-arm zones in
the images of all the bands, because it was the reddest band available
in all three galaxies, and allowed the clearest definition, minimizing
the effects of patchy dust, and of local star formation.  In Fig. 1 we
show the mask separating arm from interarm images for M~51, in Fig. 2
we show the mask for NGC~3631, and in Fig. 3 we show the mask for 
NGC~4321, in all cases overlaid on the $B$-band image of the galaxy.

\begin{figure*}
\caption{{\bf Plate 1:}
Composite plates produced by combining the $B$,$V$ and $I$ images of each of the
three galaxies: Upper image (larger) NGC~4321, Lower left image M~51 (using
$B$, $R$ and $K$),
Lower left image NGC~3631. These representations are effective in bringing out
the distribution of dust in lanes associated especially with the spiral arms
(dusky reddish features),as well as the blue star-forming zones.
}
\end{figure*}
\addtocounter{figure}{-1}
\begin{figure}
\caption{
Contour representation of the photometric map of M~51 in the
$B$-band. The boundary between the arms and the interarm region 
is shown as a thick contour (the criterion for this boundary 
is explained in the text).}
\end{figure}
\begin{figure}
\caption{As Fig. 1, now for NGC~3631}
\end{figure}
\begin{figure}
\caption{As Fig. 1, now for NGC~4321}
\end{figure}

We used the three images (total disk, arm, inter-arm) in each band for
each galaxy as inputs for the profile-fitting program Galphot (see J\o
rgensen \etal 1992). The program yields azimuthally averaged surface
brightness profiles by fitting ellipses to a two-dimensional
brightness distribution.  For each galaxy surface profiles were
calculated on elliptical isophotes. The ellipticity and position angle of
these isophotes was constant for NGC~4321 (resp. 0.37 and 25$^{\rm o}$)
and NGC~3631 (resp. 0.06 and 25$^{\rm o}$).
For M~51 we let the ellipticity increase linearly from 0.1 at the center 
to 0.25 at r = 50\sec\, and kept it at 0.25 for larger radii. Here
the position angle was fixed at 140$^{\rm o}$.
These values were determined on the basis of 
average photometry of the whole disk determined during an initial fit in $I$,
where the ellipticity and position angle were left free. 
 
The same isophotal geometry was employed for all three images of each
galaxy in a given band.  Finally radial surface brightness profiles
were produced for the whole galaxy, for the arms alone, and for the
interarm disks separately. These profiles were used to fit radial
scale-lengths to the disk in the various bands.  
Since the bulges in all of these three
galaxies are small compared to the disks, no bulge-disk decomposition
procedure was adopted, but the inner limiting radius for measuring
each disk profile was taken well outside the bulge.  The same inner
and outer radial limits were taken in all wavelengths when making the
least squares fits to derive the scale-lengths for a given galaxy, as
indicated explicitly in Table 1.
Also given in Table~1 are the formal errors from the least squares
fits. For a given range these errors are realistic. However, scale-lengths
can vary considerably with adopted range, as will later be 
shown for NGC~4321, and can vary by factors up to 2. Fortunately,
scale-length {\sl ratios} are much less range-dependent.

\begin{table}[h]
\begin{center}
\caption{Observationally derived scale-lengths, in the 
photometric bands indicated, for the arm, interarm, and whole
disk images of each of our three galaxies. Radial ranges for the
measurements are included.}
\begin{tabular}{lcrrrrrr}    
\hline
\hline
Galaxy & Phot.  Band & \multicolumn{6}{c}{Scale length  (\sec) } \\
 & & Total & $\pm$ & Arm & $\pm$ & Interarm & $\pm$ \\
\hline
M 51            & $B$ & 108. 1 & 14.9 & 154. 2  & 26.8 & 92. 8 & 9.4 \\
	        & $R$ & 91. 2  & 8.9 &  118. 0  & 11.7 & 81. 2 & 7.0 \\
Range           & $I$ & 88. 5  & 7.4 &  110. 1  & 8.5  & 80. 4 & 6.1 \\
45\sec-250\sec\ & $K$ & 87. 4  & 5.7 & 98. 7    & 7.1 &  87. 1 & 6.8 \\
\hline
NGC~3631        & $B$ & 53. 2 & 5.7 & 67. 3 & 5.7 & 49. 1 & 3.6 \\
	        & $V$ & 49. 4 & 3.9 & 58. 5 & 3.5 & 46. 5 & 2.6 \\
Range	        & $R$ & 47. 6 & 3.4 & 54. 0 & 3.1 & 45. 1 & 2.3 \\
21\sec-110\sec\ & $I$ & 41. 6 & 2.6 & 47. 8 & 2.3 & 39. 2 & 1.8 \\
\hline
NGC~4321   & $B$ & 70. 4 & 2.2 & 76. 3 & 2.8 & 60. 3 & 2.3 \\
           & $V$ & 74. 4 & 1.5 & 80. 8 & 2.7 & 56. 0 & 2.2 \\
Range      & $R$ & 76. 4 & 1.3 & 84. 5 & 2.9 & 58. 9 & 2.0 \\
35\sec-315\sec\ & $I$ & 67. 5 & 1.1 & 74. 5 & 2.3 & 58. 5 & 1.5 \\
\hline
NGC~4321        & $B$ & 155.0 & 18.8 & 195.1 & 64.8 & 93.4 & 7.4 \\
                & $V$ & 109.9 & 6.6  & 137.7 & 18.0 & 76.4 & 4.6  \\
Range           & $R$ & 100.4 & 5.7 &  124.8 & 11.9 & 72.2 & 4.2 \\
35\sec-100\sec\ & $I$ & 86.3  & 4.2 &  106.5 &  5.8 & 65.8 & 3.5 \\
\hline
NGC~4321         & $B$ & 61.2 & 1.9 & 72.6 & 4.7 & 49.8 & 1.9 \\
                 & $V$ & 69.1 & 2.2 & 80.9 & 5.6 & 46.3 & 2.2 \\
Range            & $R$ & 72.7 & 2.3 & 87.8 & 6.6 & 50.3 & 2.4 \\
100\sec-315\sec\ & $I$ & 63.5 & 1.7 & 76.0 & 5.0 & 51.3 & 1.7 \\
\hline
\end{tabular}
\end{center}
\end{table}

\section{Observational results}

\subsection{M~51}

In Fig.  4 (a) we show azimuthally averaged radial surface brightness
profiles for the whole galaxy in the $B$, $R$, $I$, and $K$ bands, in
Fig.  4 (b) the corresponding  profiles for the arm regions of the disk,
and in Fig. 4 (c) the profiles for the interarm zones.  
The best fit
to the exponential disk component is indicated in each figure.  The
fitted values of the observed disk scale-length 
are listed in
Table 1 for each photometric band, as well as the limits of the
radial ranges over which the fits were obtained.
If we look at the
scale-lengths for the disk as a whole, we see a modest change between
the $B$ and the $K$ bands: a fall of some 25\%, but this masks a major
difference between the arms and the inter-arm zones. In the arms the
$B$ scale-length is longer than that in K by a factor of 1.6, whereas
in the interarm disk the scale-lengths change very little over the
whole wavelength range. This stronger trend of scale-length variation
with wavelength in the arms almost certainly results from more dust
there than in the inter-arm zones, and we will examine this below using our
theoretical models for comparison. The whole disk scale-lengths are
quite close to those in the interarm disk, which means that for M~51
the optical effect of the arms, in the bands observed, does not greatly
distort the average scale-lengths.

\begin{figure}
\centerline{\epsfysize=7cm \epsfbox{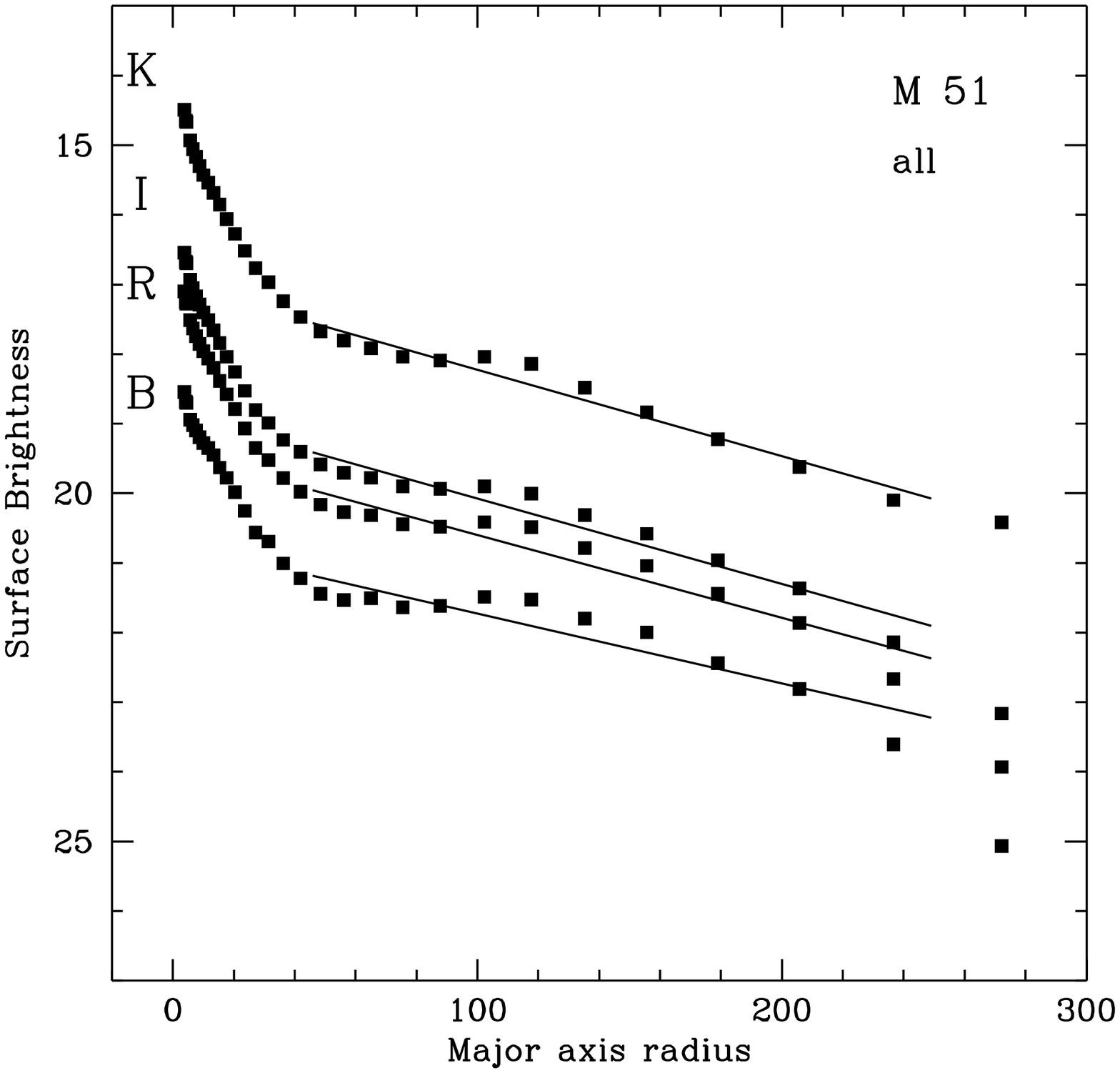}}
\centerline{\epsfysize=7cm \epsfbox{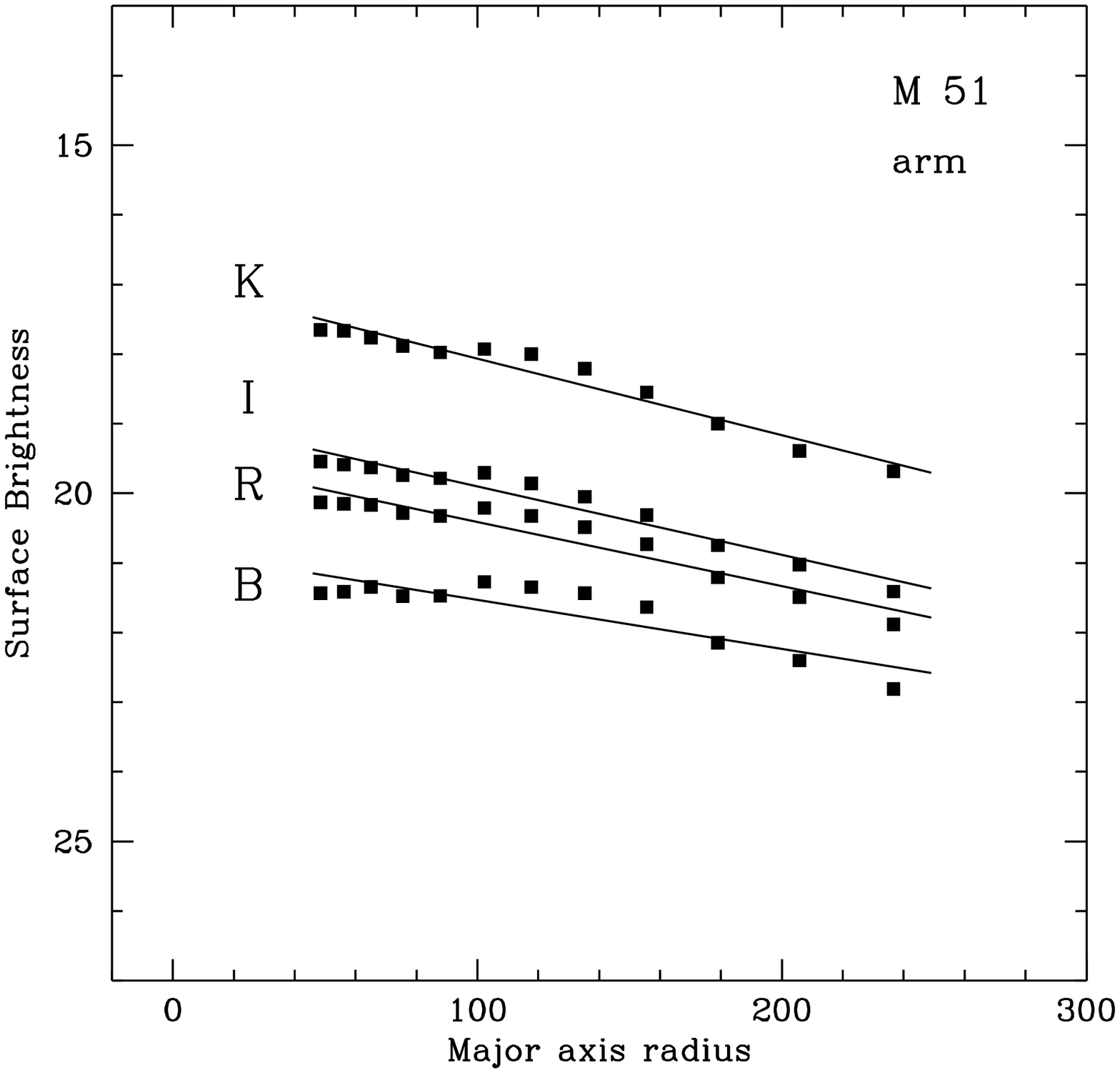}}
\centerline{\epsfysize=7cm \epsfbox{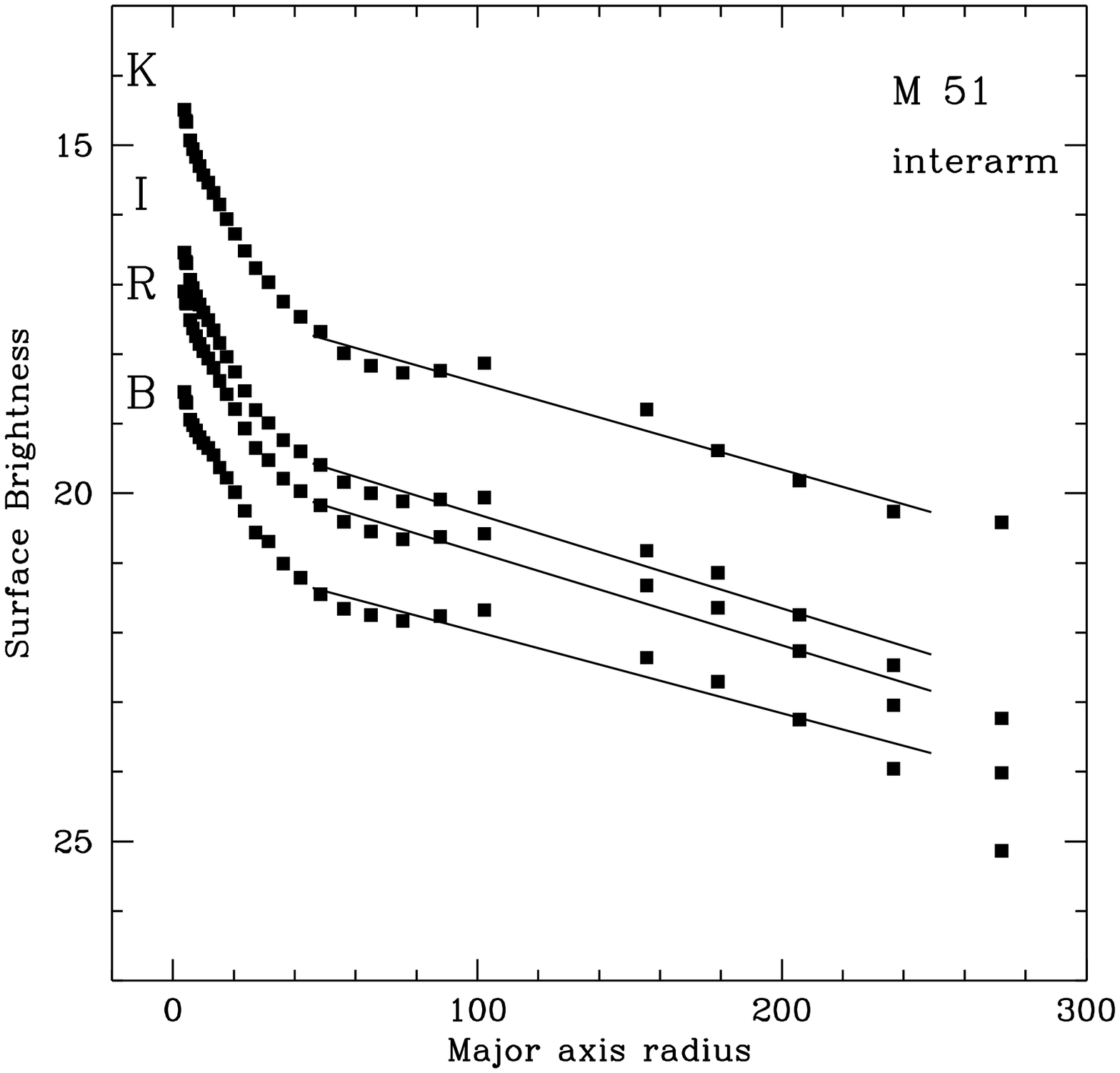}}
\caption{
Radial surface brightness profiles for M~51 for the whole
galaxy (a), the arms (b), and the interarm region (c).
Also plotted are the exponential fits to the surface brightness
profiles that best fit in the range over which the lines have been drawn.}
\end{figure}

\subsection{NGC~3631}

In Fig. 5 (a) we present the radial surface brightness profiles for
this object in the $B$, $V$, $R$, and $I$ bands, in Fig. 5 (b) we show
the profiles for the arms only, and in Fig. 5 (c) for the interarm
zones.  The fitted values for the observed scale-lengths and the
extrapolated on-axis surface brightnesses are given in Table 1.
The scale-length behavior of NGC~3631 is qualitatively similar to
that of M~51.  In the arms there is a systematic decline of
scale-length with increasing wavelength, which is barely evident in
the interarm disk. The scale-lengths in the arms are systematically
longer than those in the interarm zones, but converge towards equality
as the wavelength increases.  We do not have K-band information here,
but in this convergence, NGC~3631 shows similarity to M~51. Here, too,
the most straightforward interpretation of the trends observed is that
in the arms the optical effects of dust extinction are readily
detectable, whereas in the inter-arm zones they are not. We will use
comparison with our models in the next section to quantify these
considerations. 
As for M~51 the average scale-lengths for the whole disk
of NGC~3631 are
closer in value to those of the interarm zones. 
The outer parts of the profiles fall away more steeply then in the zone
whose scale-lengths we have measured, but the signal to noise ratios
are not high here, so scale-length measurements here would not
be reliable, although we can obtain a rough indication
of uniformity with wavelength.

\begin{figure}
\centerline{\epsfysize=7cm \epsfbox{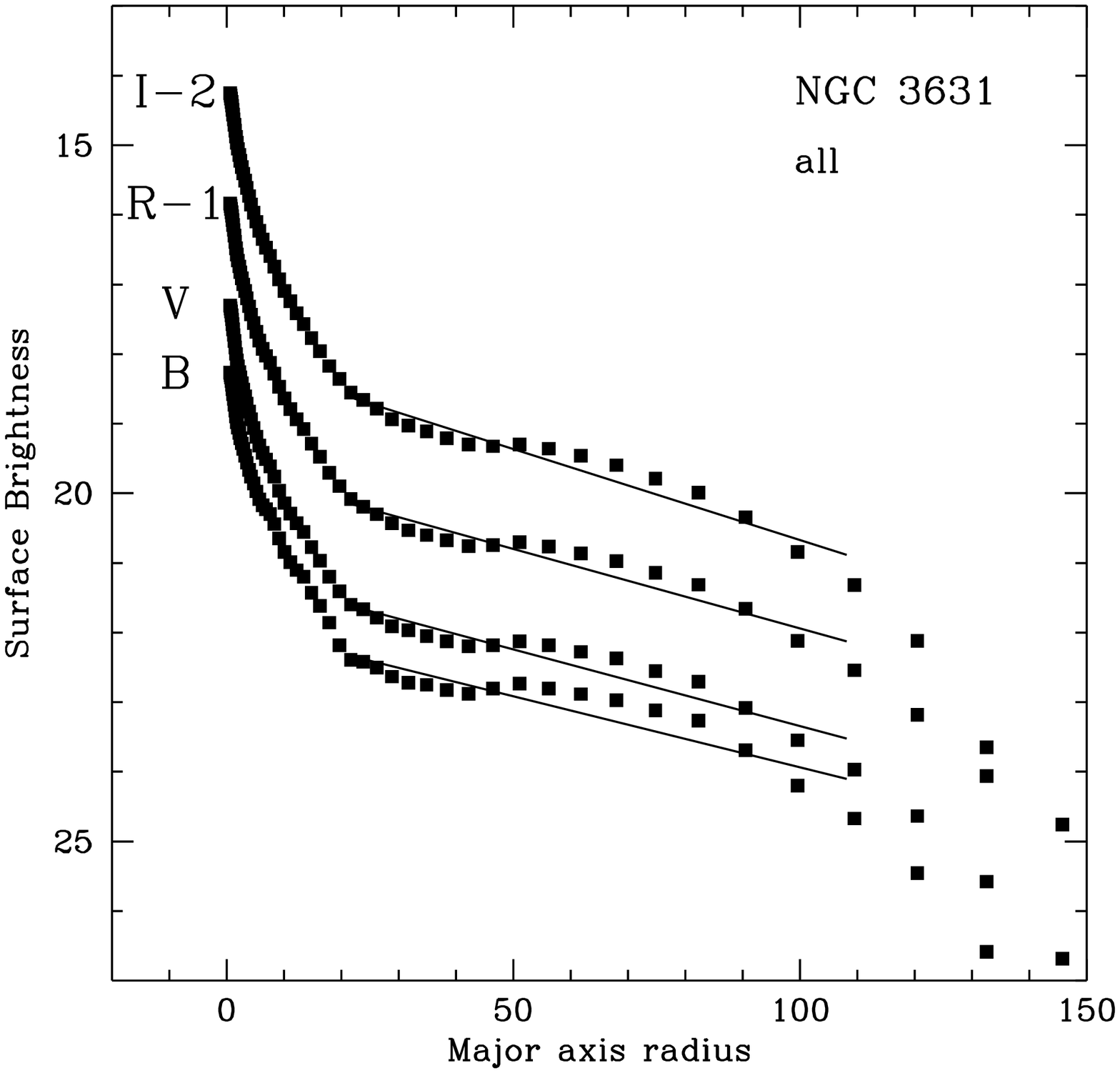}}
\centerline{\epsfysize=7cm \epsfbox{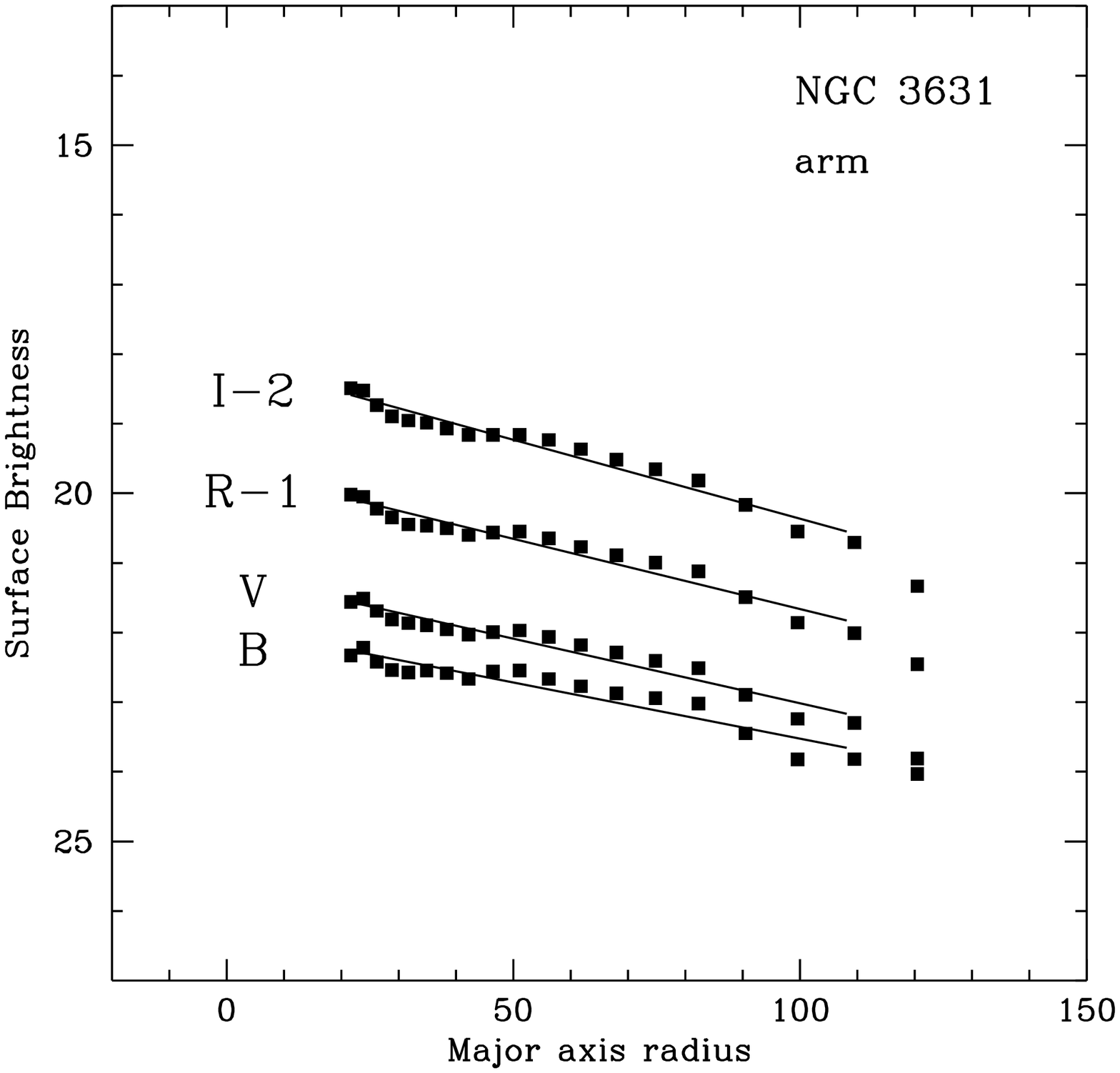}}
\centerline{\epsfysize=7cm \epsfbox{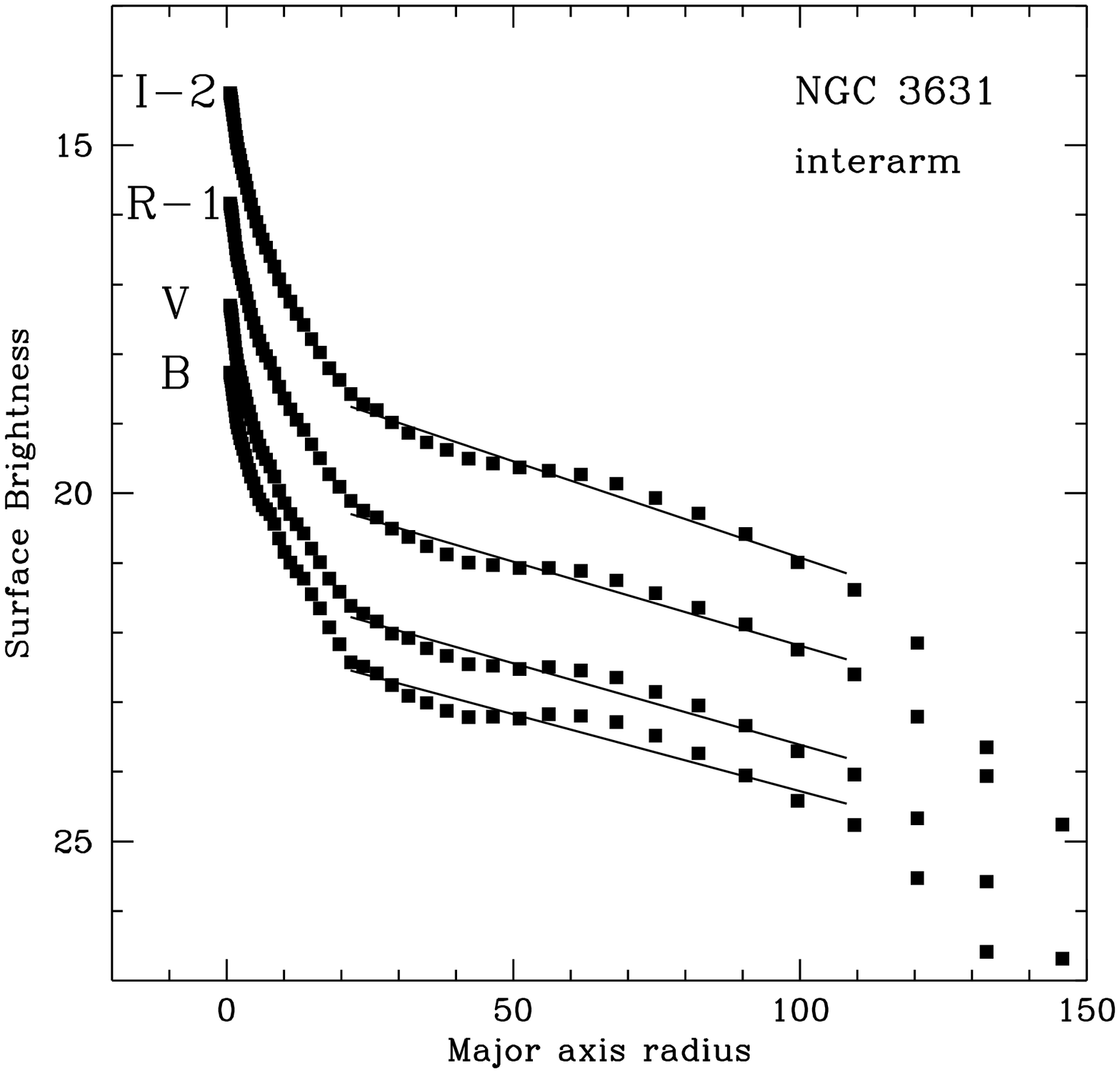}}
\caption{
Radial surface brightness profiles for NGC~3631 for the whole
galaxy (a), the arms (b), and the interarm region (c).
Also plotted are the exponential fits to the surface brightness
profiles that best fit in the range over which the lines have been drawn.}
\end{figure}

\subsection{NGC~4321}

Figure 6 shows the radial photometric profiles of this galaxy: in
Fig. 6 (a) the averages for the whole disk, in Fig. 6 (b) the profiles
for the arms alone, and in Fig. 6 (c) those for the interarm
zones. The fitted scale-lengths are again given in Table~1. There
are notable differences in profile behavior between NGC~4321 and the other 
galaxies. In NGC~4321 the profiles divide clearly into two radial
zones, from 35'' to 100'' from the center, and from 100'' to 
315''. The inner zone corresponds to the bar reported, among others,
by  Pierce (1986)
and studied photometrically and kinematically in Knapen \etal (1993),
and the outer zone compares to the disk beyond the bar. We should note that
our observations here go out to some 5 scale-lengths from the center,
compared with only 3 scale-lengths for the other two galaxies.

\begin{figure}
\centerline{\epsfysize=7cm \epsfbox{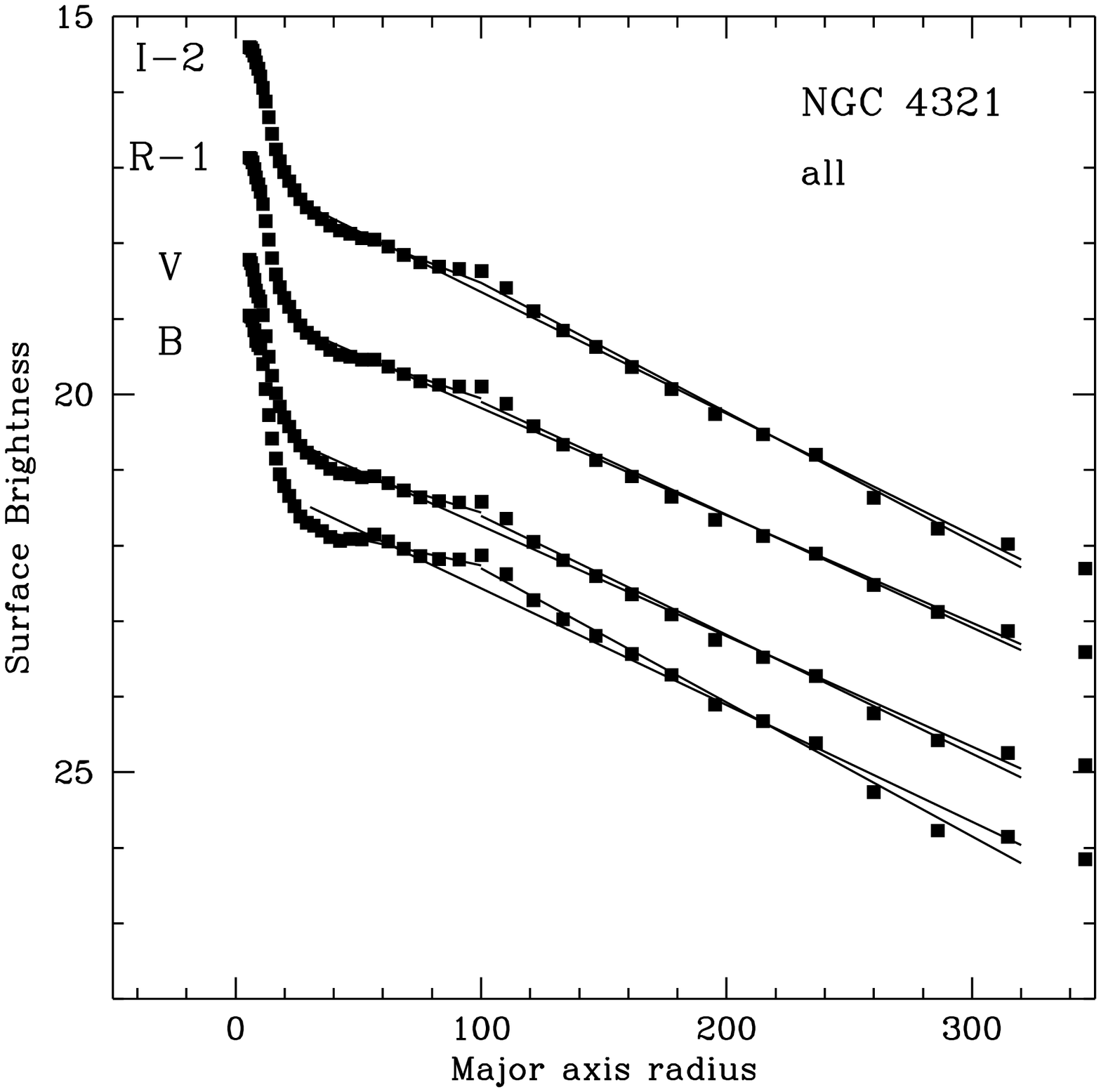}}
\centerline{\epsfysize=7cm \epsfbox{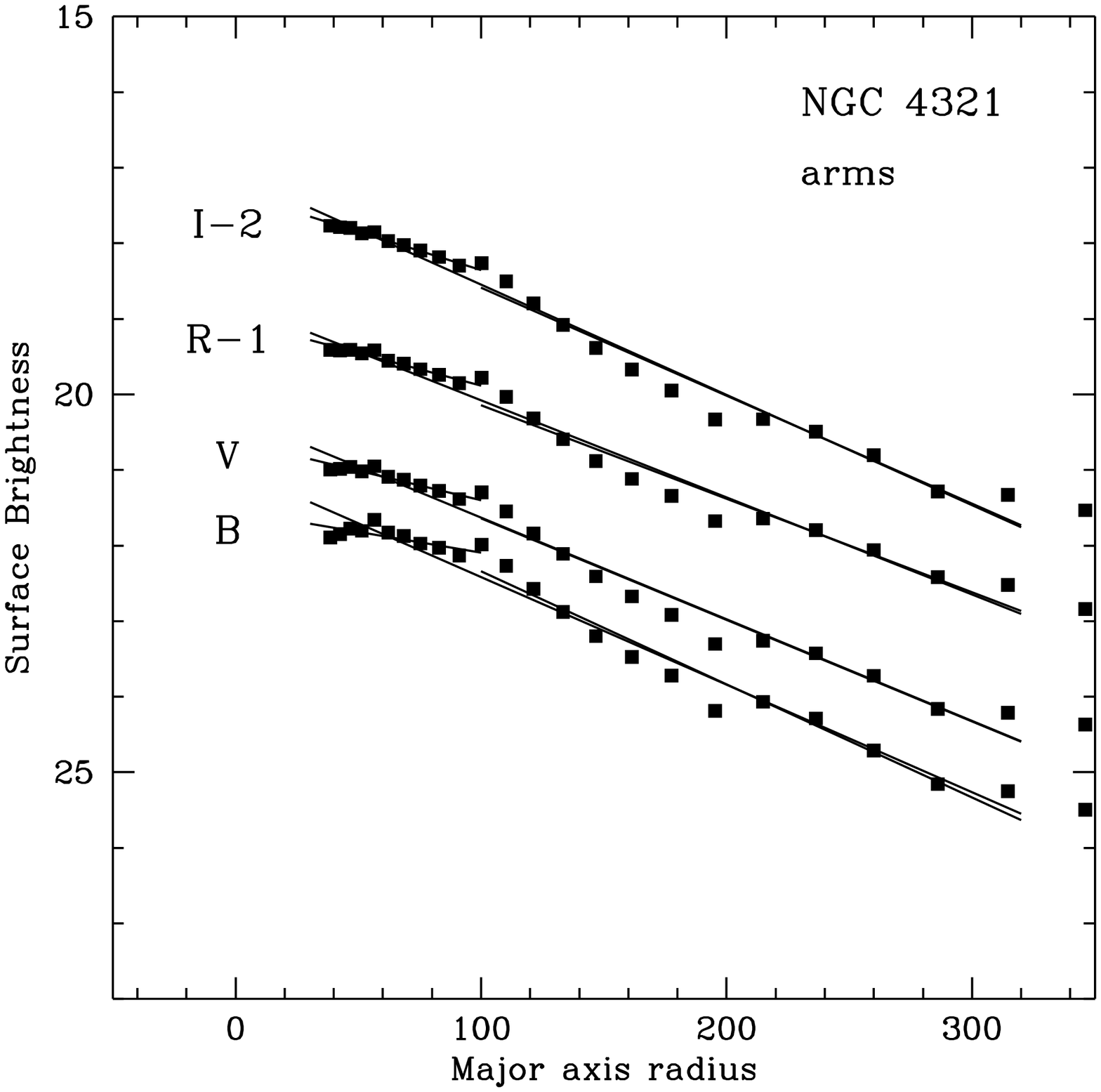}}
\centerline{\epsfysize=7cm \epsfbox{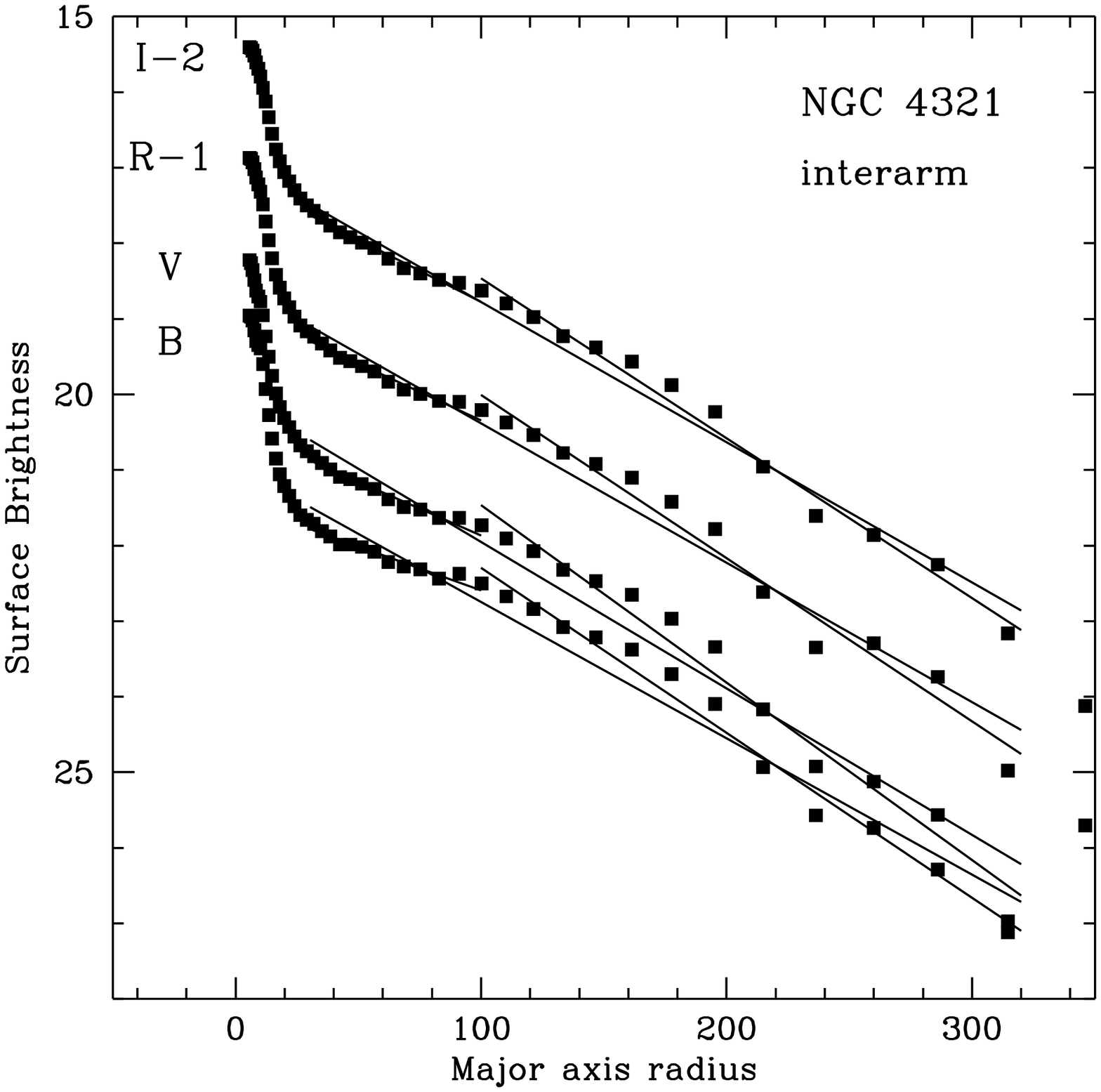}}
\caption{
Radial surface brightness profiles for NGC~4321 for the whole
galaxy (a), the arms (b), and the interarm region (c).
Also plotted are the exponential fits to the surface brightness
profiles that best fit in the range over which the lines have been drawn.}
\end{figure}

The behavior of the inner zone is quantitatively similar to that
of the disks of the other two objects. Both arm and interarm scale-lengths 
decline with increasing wavelength, as expected from systems containing
measurable quantities of dust. The 'arm' scale-length is larger,
and appears to be converging to a longer value than the interarm lengths,
implying a slower fall-off in intrinsic stellar surface brightness
in the arms (by no means unexpected). 

The behavior of the outer zone is qualitatively different. Here too
the arms show considerably longer scale-lengths, showing that the 
arm-interarm contrast is rising with increasing radius; this must be a 
property of the surface density of the stellar population. 
However there are no systematic trends to scale-lengths at longer
wavelengths in this radial range. The interarm zone shows virtually no
variation at all, while the arms show a slight increase from $B$ to
$R$, followed by a slight dip at $I$. In neither case is the behavior
readily consistent with major quantities of dust, and we can conclude that
there is relatively little dust in the outer parts of the disk of 
NGC~4321.

\section{Comparison of observations with models:  the role of dust}

\subsection{Tests for dust in spirals}

There has been considerable discussion in the recent literature about
the possible presence of major quantities of dust in galactic disks,
and of the implications for the measured properties of galaxies, in
particular for their mass to light (M/L) ratios.  A variety of
observational tests has been brought to bear on the problem.
Valentijn (1990) used the variation of surface brightness with
inclination angle for a large sample of galaxies taken from the ESO-LV
catalog (Lauberts \& Valentijn 1989), reaching the conclusion that
disks must be optically thick in some agent causing extinction (not
necessarily dust), because the surface brightness does not increase
sufficiently with increasing inclination angle for an optically thin
system.  Valentijn's conclusions challenged the classical study by
Holmberg (1958), who used  the same test, and  
claimed clear evidence for
low optical thickness in disks.  

Absolute measurements of dust extinction have recently been
published for a small number of nearly edge-on galaxies (Knapen
\etal 1991; Jansen \etal 1994;   
Byun \etal 1994),  with values of peak
optical depths of between 2 and 3 in the central,  densest parts of
the dust lanes observed. Assuming that this dust is distributed in an
exponential form,  comparable to the stars or gas  in a typical
galactic disk, we would infer that the same galaxies,  when observed
face on,  would exhibit optical depths in $B$ or $V$ well below unity
over all but possibly the innermost parts of their disks.  If on the
other hand the dust is concentrated into narrow lanes the face-on
optical depth of a lane might well be significantly higher than this
but the optical depth averaged over the disk is very unlikely to be so. 

In a study of the surface photometry in $B$ and $K$ of 37 Sb's and
Sc's, Peletier \etal  (1995) found that the $B-K$  colors in the centers
of the disks of spirals are generally more than 1 magnitude redder
that at D$_{25}$.  This, and the inclination dependence, implies that
the extinction in the $B$ band in the centers of face-on galaxies is
generally of order 1 magnitude.  However the dust morphology in
galaxies is not monotonically exponential, as we have seen for the
galaxies treated in the present paper, and the absolute extinction in
edge-on objects could well be attributable to single dust lanes.  If
we assume this, and take the geometrical width of a dust-lane to be of
the order of half the width of a spiral arm (see e.g. Plate 1), we
would infer from the edge-on galaxy measurements that face-on
dust-lane depths are indeed of order 2 or 3.  We note again, however,
that the optical depths over the rest of the disk are likely to be
much lower.  

Kinematic tests, using the wavelength dependence of the rotation
curves of edge-on galaxies have been performed by Bosma
\etal (1992), and by Prada \etal  (1994, 1996). 
The former authors,  combining
radio and optical data, inferred low dust optical depths, at least in
the outer parts of the Sc spiral NGC~100, and the Sb spiral NGC~891.
The latter authors, using optical and NIR spectroscopy, found an
optical depth of $\sim$ 30 in the SABc galaxy NGC~253 and 
close to 10 along a line of sight close to the center of
the Sb pec.  galaxy NGC~2146  These values  would translate to face-on values
of resp. 3 and 1 for a homogeneous exponential dust distribution.
Another test, which is revealing, but almost impossible to
apply on a statistical basis, was used by Andredakis and van der Kruit
(1992), who studied the colors of a galaxy seen through the outer
parts of a disk of a nearer Scd 
galaxy, concluding that the foreground
spiral was optically thin at the point of observation.  In a similar
observation White \& Keel (1992) studied a galaxy seen behind a
foreground Sbc galaxy and concluded that the nearer object exhibited
significantly more extinction in the arms than between them.
These measurements are consistent with the view of
Valentijn (1990) that galaxies of type 3-5 (Sb-Sc) contain significantly
more dust than Scd galaxies (type 6).

The differential scale-length test used in the present paper is
potentially powerful, as it is susceptible to a statistical approach.
In a disk with a uniformly distributed stellar population falling
exponentially in surface density the observed scale-length will vary
with wavelength, if there is a significant amount of dust.  For dust
obeying the same extinction law as in our Galaxy (see Knapen \etal
1992; and Jansen \etal 1994 for evidence that this is a reasonable
assumption to take) the observed scale-length will decrease
systematically with increasing wavelength in the presence of
significant quantities of dust, also distributed exponentially. To
illustrate the magnitude of the effect with an example, if the
intrinsic radial surface density distributions of the stellar
population and the dust follow the same exponential trend and the dust
has a scale-height above the mid-plane of the galaxy one half that of
the stars, the observed scale-length in the $B$ band will be twice
that in the $K$ band for an on-axis extrapolated optical depth of 3 in
B.  The ratio of the $B$ to $I$ scale-lengths in this case would be
1.4 (see Peletier \etal (1995) for more examples).  
Differences of this order are not especially difficult to
measure with the required accuracy, over the first 3 scale-lengths
of a disk, if the area covered by the detector is large enough.
Previous studies
in which the scale-length as a function of wavelength has been
presented, or whose radial color variations have been examined,
include that by Elmegreen and Elmegreen (1984), Rix and Rieke
(1993), Evans (1994), Peletier \etal (1994) and de Jong (1995).  
Of these, Rix and Rieke,
Evans, Peletier \etal\ and de 
Jong offer interpretations explicitly in terms of dust
extinction. In fact Evans drew the negative conclusion that the
measurements available to him were not of sufficient quality to
distinguish between major and minor quantities of extinction in dust.
However, the importance of the morphological detail: the quantitative
separation of arm and inter-arm disk, and the importance of the
geometry of the structures in which the dust is distributed have
previously been discussed only briefly by the present authors (Beckman
\etal 1995) when considering large-scale extinction by dust in disk
galaxies. 
 
\subsection{The Model}

We have developed our own models for the transfer of
radiation through disk galaxies, which take into account the effects
of absorption and of multiple (diffuse) scattering by dust. In this
respect they differ from the ``triplex" models of Disney \etal (1989),
although the basic scheme and underlying concept is owed to these
authors. The radiative transfer equation was solved for a local plane
parallel geometry, using a moment method with a high order closure
relation for the moment system, in order to take into account the
angular phase function of the dust. The moment system was solved using
an implicit method which assures unconditional stability (Golulo \&
Ortega 1992). The algorithm will be presented in a forthcoming paper
(Corradi, Beckman \& Simonneau 1996, in preparation). 
The scattering phase function
of Henyey and Greenstein (1941) was adopted throughout, and the dust
parameters: opacity, albedo, and the scattering anisotropy parameter,
were taken from Di Bartolomeo \etal  (1995).  These models were used
as first order approximations to the disks of galaxies observed
face-on ($i=0$).  Axisymmetric exponential distributions of stars and
dust are assumed, in two perpendicular dimensions which represent the
radial and axial directions, $R$ and $z$, respectively, as expressed in
equations (1) and (2):
 
$$\epsilon (R, z)  ~=~ \epsilon (0) ~\exp ^{(-R/R_s - z/z_s)}  ~~~~~~ (1)  $$
 
$$\alpha (R, z)  ~=~ \alpha (0) ~\exp ^{(-R/R_d - z/z_d)}  ~~~~~~ (2)  $$

where $(R, z)$ are cylindrical coordinates, $\epsilon (R, z) $ is the
volume emissivity of the stellar contributors, $\alpha (R, z) $ is the
dust opacity, and $R_s$, $z_s$ and $R_d$, $z_d$ are the scale-lengths
and the scale-heights of the stars and the dust, respectively.  In
order to limit the number of free parameters, we have assumed in the
present exercise that the stars and the dust have the same radial
scale-lengths i. e.  that $R_s$ = $R_d$. These one-dimensional models
describe correctly the exponential variations of the volume emissivity
and the opacity in the direction of the z-axis of a disk, and variable
values can be incorporated for the stellar to dust scale-height
ratio. The plane-parallel geometry does not, of course permit a
perfect representation of the radial exponential decline in the disk.
To account for this we make the approximation that at a given radius
$R$ the galaxy can be treated as a plane-parallel system,
with $\epsilon (R, z) $ and $\alpha (R, z) $ given by eqs.  (1) and
(2).  This approximation is valid, given the large ratios of
scale-length to scale-height in the disks of galaxies, if the mean
free path of a photon is short compared with the scale of the
variation of the volume emissivity.  This condition does not strictly
hold for low optical thicknesses, i.e.  at large radii or for large
heights above the planes of galaxies, but the present models
nevertheless can give valuable insight into the effects of dust
extinction (and in particular of the effect of scattering) on galaxy
photometry. 

We will show here results obtained using two modes of the
models: one in which the scattering albedo was set to zero, i. e.
with absorption only, and the full model with the same 
opacity but with realistic scattering parameters included. We note
that in the first case the models in fact yield the exact solution of
the radiative transfer equation for a face-on system.  The models run
for comparison with the observations presented in the present paper
were for a range of galactocentric distances out to 5 intrinsic
scale-lengths, and used two ratios of stellar to dust scale-height
ratios: 1 and 3, to comply with a range of observationally observed
dust distributions seen in edge-on systems (see Jansen \etal
1994). The models were parametrized in terms of their central optical
depth (i. e.  the optical depth extrapolated to the central axis of
the disk) in the $V$ band $\tau_V$ (= 2$\alpha (0) ~z_d$) and the
corresponding values of $\tau$ for the other photometric bands were
scaled from the dust opacities in Di Bartolomeo \etal (1995) based on
the Galactic extinction law (Rieke \& Lebofsky 1985).
The models were then run for the bands $B$ $V$ $R$ $I$ and
$K$ according to the observational data to be fitted for each galaxy.
The direct outputs from the models are a set of theoretical radial
luminosity profiles for the dusty disks.  The associated scale-lengths
are then computed by taking a linear fit to the logarithmic
representation of the intensity of the profile between selected inner
and outer radii.  In general a disk with a truly exponential
distribution of stars and dust does not yield an exponential intensity
profile, but the deviations are not large, and decrease with
increasing radial range.  In consequence there will be a 
dependence of the derived scale-length on the values chosen for the
radial limits in a given model.  Here we have adopted the simple
procedure of choosing, for a given galaxy, the same values for the
radial limits in our models as those used in practice for deriving the
observed scale-lengths.  These limits are noted in Table 1.
 
\subsection{The model fits}

\subsubsection{M~51}

In Fig.  7 we present 
the results of fitting the models described in Section 4.2
to the data in Table 1 for the case of M~51.  The 4 model fits shown
are: 7 (a) and 7 (b) for dust with absorption only, no scattering, and
with two different ratios of stellar to dust scale-heights: 1 and 3
respectively; 7 (c) and 7 (d) incorporate scattering for the same two
ratios of stellar to dust scale-heights.  In all 4 diagrams the
observational data are shown in the form of ratios, normalized to the
scale-length of the galaxy as a whole at the longest waveband used (K
band). The wavelength dependence, stronger in the arms than in the
interarm zone, is readily seen, and the values for the whole disk lie
closer to those for the interarm zone. For comparison we include in
each diagram the theoretical curves for 5 values of the central,
visual optical depth $\tau_V$ (extrapolated to the central axis) of
the galaxy. The values selected for $\tau_V$ are 1, 2, 4, 10 and
20. These
comparisons yield the qualitative conclusion that the optical
depth in M~51 has moderate on-axis values. In Fig 7 (d), the case
favoring most dust, the optical depth in $V$ in the arms, extrapolated to
the axis, takes a value of 10, 
and that in the interarm disk a value close to 
1, and
that of the galaxy as a whole a value between  2 and 4.  This assumes that the
scale-height in dust is only one third that of the stars, which is a
conservative assumption as far as the young stellar component is
concerned, and leads to a slight over-estimate 
of dust absorption. The
value of the optical depth averaged out to a radius of 3 scale-lengths
from the center would be 0.7. These values in fact give an
intuitively false, exaggerated idea of the effective light absorption,
because we are used to situations - those within our Galaxy- where the
star or stars observed are all behind the absorbing dust, a so-called
``screen" situation. In fact, in the model represented in Fig 7 (d) 
for the whole disk the
fraction of the light in $V$ intercepted by dust, averaged out to a
radius of 3 scale-lengths, is only 12\%, and that is the relevant
number to bear in mind when asking whether or not the dust in M~51
could lead to a bad underestimate of the total stellar content.
In fact the model stellar to dust scale-height ratio of 1 (Fig. 7c)
gives better looking fits to the data than that for ratio 3 (Fig. 7d),
but we have used the numbers for the latter case as being more conservative
in giving a greater amount of dust extinction.

\begin{figure}
\centerline{\epsfysize=13cm \epsfbox{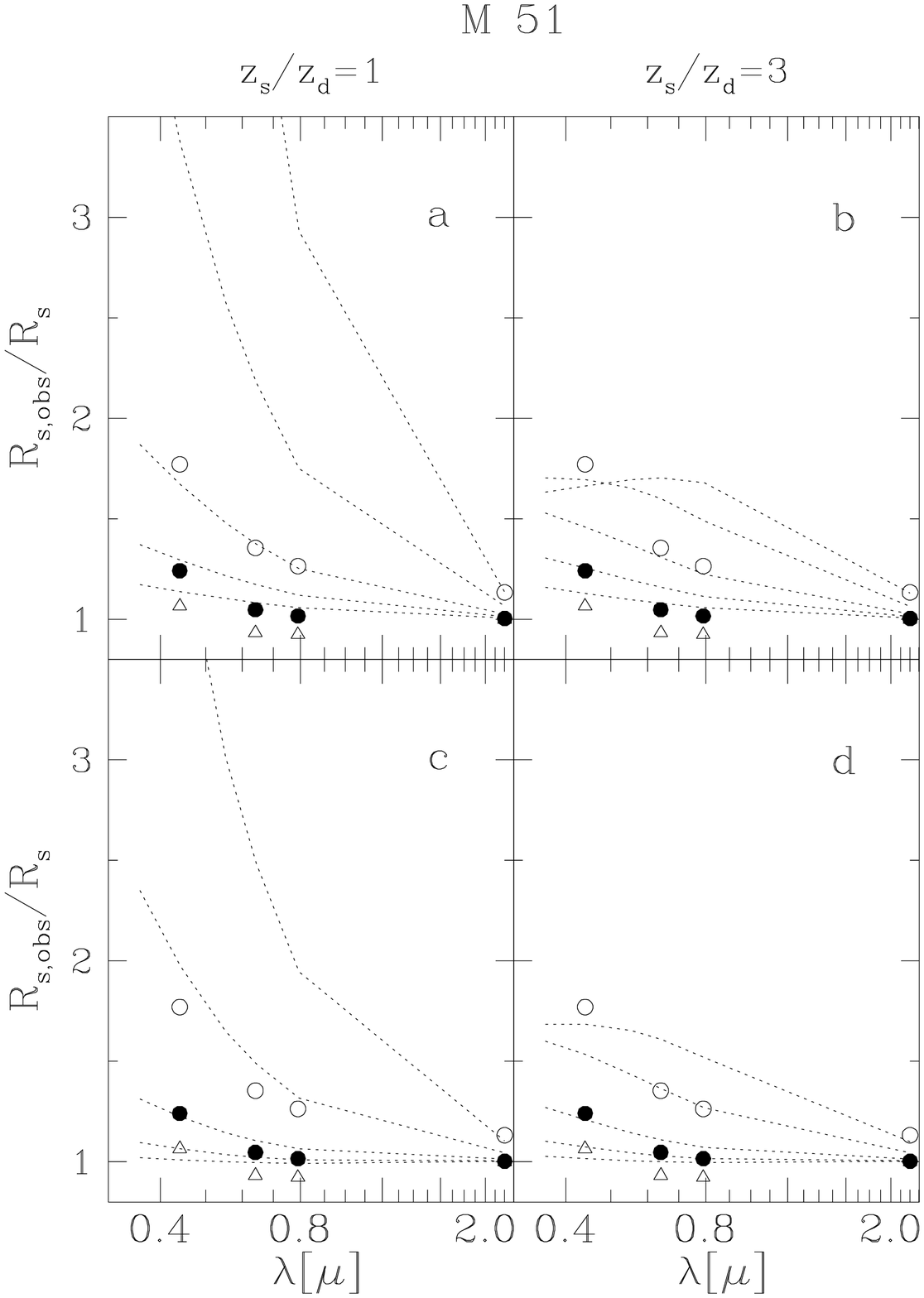}}
\caption{
Scattering models of M~51. Plotted are scale-length
ratios for the interarm region (open triangles), the arms (open circles)
and the whole disk (filled circles), together with model curves for
$\tau_V$ = 1, 2, 4, 10 and 20. The upper models include absorption only,
while the lower models include absorption and scattering.} 
\end{figure}
 
\subsubsection{NGC~3631}

The model fits to the data for NGC~3631 are shown in Fig. 8.
Equivalent cases are dealt with to those shown for M~51, i. e.
absorption alone, and absorption plus scattering, for the two ratios,
1 and 3,  of the stellar to dust scale-height ratio.  Here again the
pure absorption cases requires a lower optical depth in dust than the
scattering cases, but we treat only the latter as serious candidates
for good model fits, since real dust must in fact show scattering.
The general behavior of the arms, the interarm zones, and the
averaged disk is quite similar, in this galaxy, to that of M~51, and the
scattering models in Figs 8 (c) and 8 (d) give very fair fits to the
observations, when the on-axis V-band optical depth for the arms takes
the value 10; for the interarm zones the corresponding value is
close to 4, and for the galaxy as a whole the best value is
6.  The average value for the dust optical depth in $V$ out to a
radius of 3 scale-lengths is 1.1 for this galaxy.  We should note
that this average (as for M~51), is taken over the disk, but
extrapolated to cover the innermost zone, occupied by the bulge, so
that the value 1.1 represents an upper limit. The bulge really ought
to be excluded from present considerations.  This would have implied
averaging between 0.3 scale-lengths and 3 scale-lengths to get a true
disk estimate.  Giving $\tau_V$ this conservatively high mean value of
1.1, however, we find using the corresponding flux integration that no
more than 20\% of the light in the disk, out to 3 scale-lengths from
the center, has been removed by dust. Just like in the case of 
M~51, the star to dust  
scale-height ratio of 1 appears to give better fits to the data than the
ratio 3; here too we have preferred to use the latter model to
quantify the dust, in order to over- rather than  underestimate the 
extinction.

\begin{figure}
\centerline{\epsfysize=13cm \epsfbox{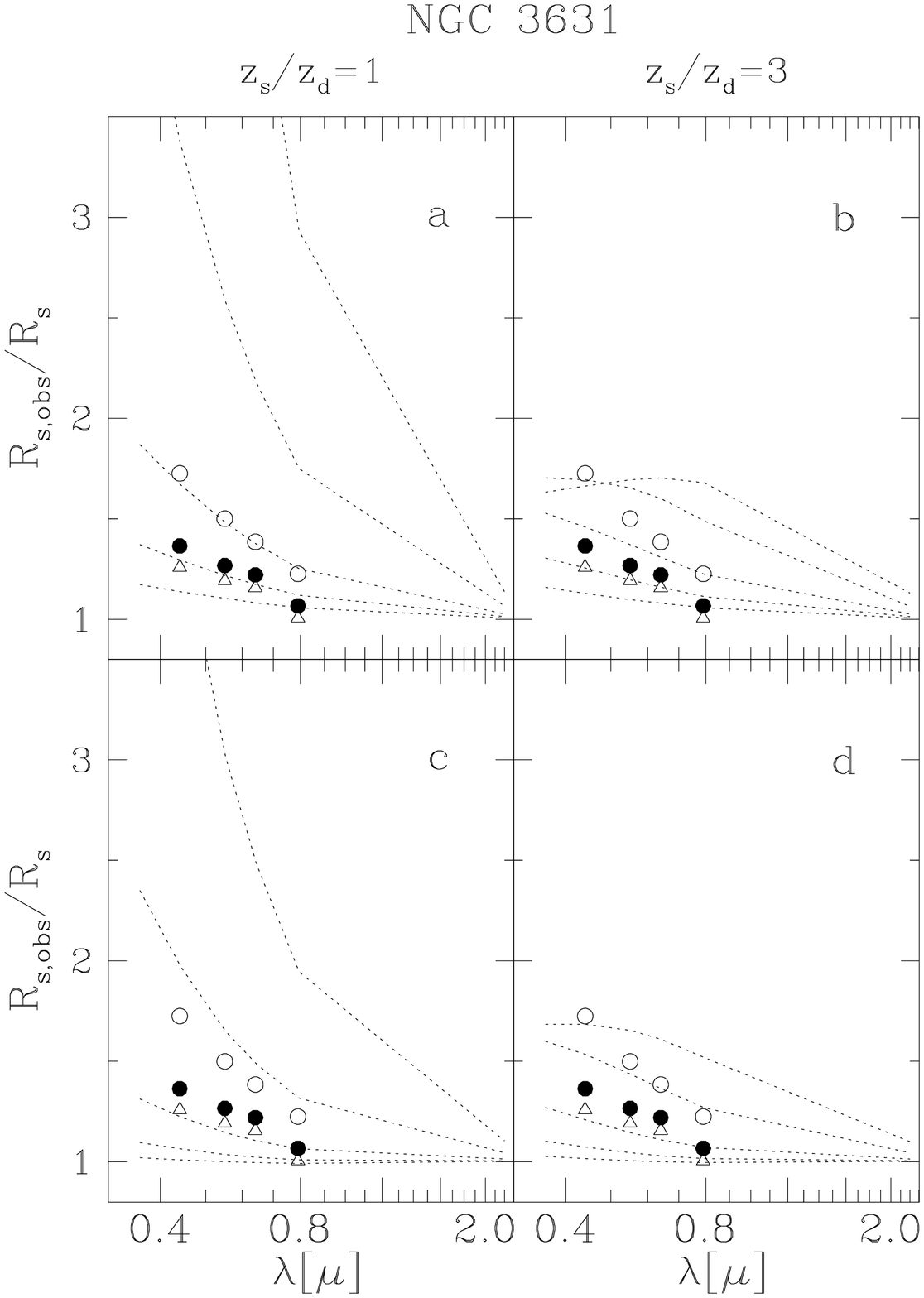}}
\caption{
Scattering models of NGC~3631. Plotted are scale-length
ratios for the interarm region (open triangles), the arms (open circles)
and the whole disk (filled circles), together with model curves for
$\tau_V$ = 1, 2, 4, 10 and 20. The upper models include absorption only,
while the lower models include absorption and scattering.} 
\end{figure}

\subsubsection{NGC~4321}

In NGC~4321 we have seen that the disk is clearly divided
into an inner zone with monotonic wavelength variations for the 
scale-lengths in the arms and the interarm regions, and an outer region,
where there is no systematic variation. In Fig.~9 we show the model fits 
for the two values of the stellar to dust scale-height ratio; 3 and 1,
and for dust with and without scattering, as applied to the inner 
zone. It is clear from these Figures that for stellar to dust
scale-height ratio 3 the models are not able to fit the data,
whereas for a ratio of unity, fits are possible. 
We opt for the scattering case, Fig. 9c, as physically more realistic,
and find that the optical depth in the arms has an on-axis extrapolated
value of 8, the equivalent for the inter-arm zone is close to 3, and for the 
averaged disk 6. The radial extent of this zone is approximately the 
same as its own scale-length, and the estimated fraction of light 
removed by dust is some 40 \%  in this zone. 

\begin{figure}
\centerline{\epsfysize=13cm \epsfbox{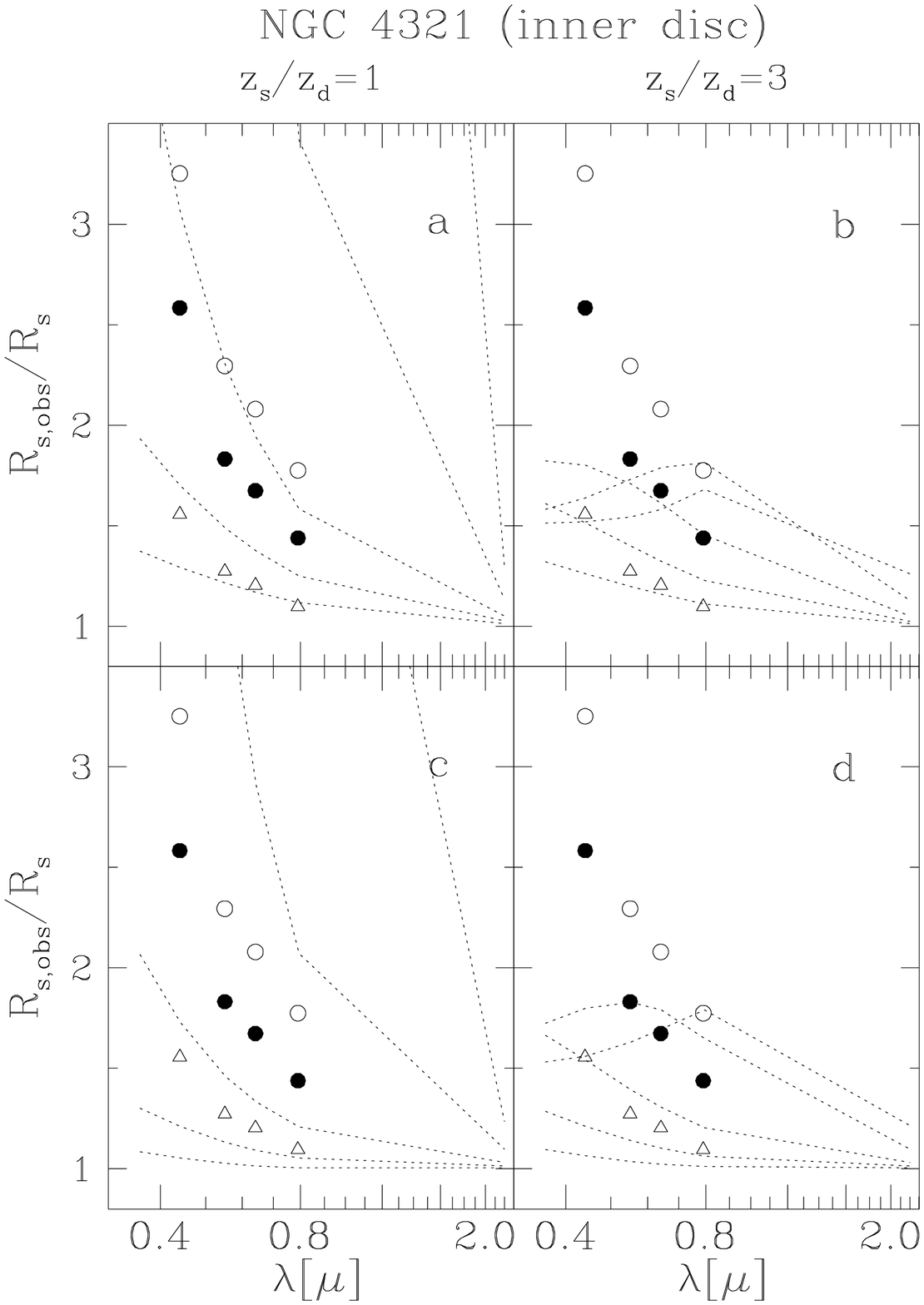}}
\caption{
Scattering models of NGC~4321 for the inner range. Plotted are scale-length
ratios for the interarm region (open triangles), the arms (open circles)
and the whole disk (filled circles), together with model curves for
$\tau_V$ = 1, 2, 4, 10 and 20. The upper models include absorption only,
while the lower models include absorption and scattering.} 
\end{figure}

The outer zone is much more difficult to quantify, as
the observed scale-lengths do not give fits to the models, showing that
one or more of the model assumptions does not hold (See Fig.~10). The 
interarm disk is characterized by essentially invariant scale-lengths,
while the arms show slightly larger scale-lengths in $V$ and $R$
than in $B$ and $I$, but all within a 20\%  range of values,
and all some 50-60\%  longer than the interarm values. Our judgement 
about these results is that the basic difference between the arm
and the interarm values is due to the fact that the arms have a stellar
population distribution with a shallower radial slope i.e. the 
arm-interarm contrast in stellar surface density rises with
increasing radius. The fine detail in $R$ and $I$ in the arm values 
is probably due to the effects of radial star formation differences
on the arm population, but this speculative and not of prime importance
for the present study, and we will not pursue it further here.
A zero order conclusion about the dust in the outer zone
is that there is very little in the interarm zone. This conclusion
is very clear, if we exclude an exact cancellation of dust reddening
and population blueing as the cause of the wavelength-invariance
of the scale-lengths. In the arms it is also a measurable
inference. The outer disk clearly dominates when we fit the whole
range of the galaxy between 35'' and 315'', as we see in Fig.~11,
so that the behavior here qualitatively is the same as in Fig.~10.
We will discuss below in detail the possible effects of dust 
in lanes, rather than distributed uniformly exponentially.

\begin{figure}
\centerline{\epsfysize=13cm \epsfbox{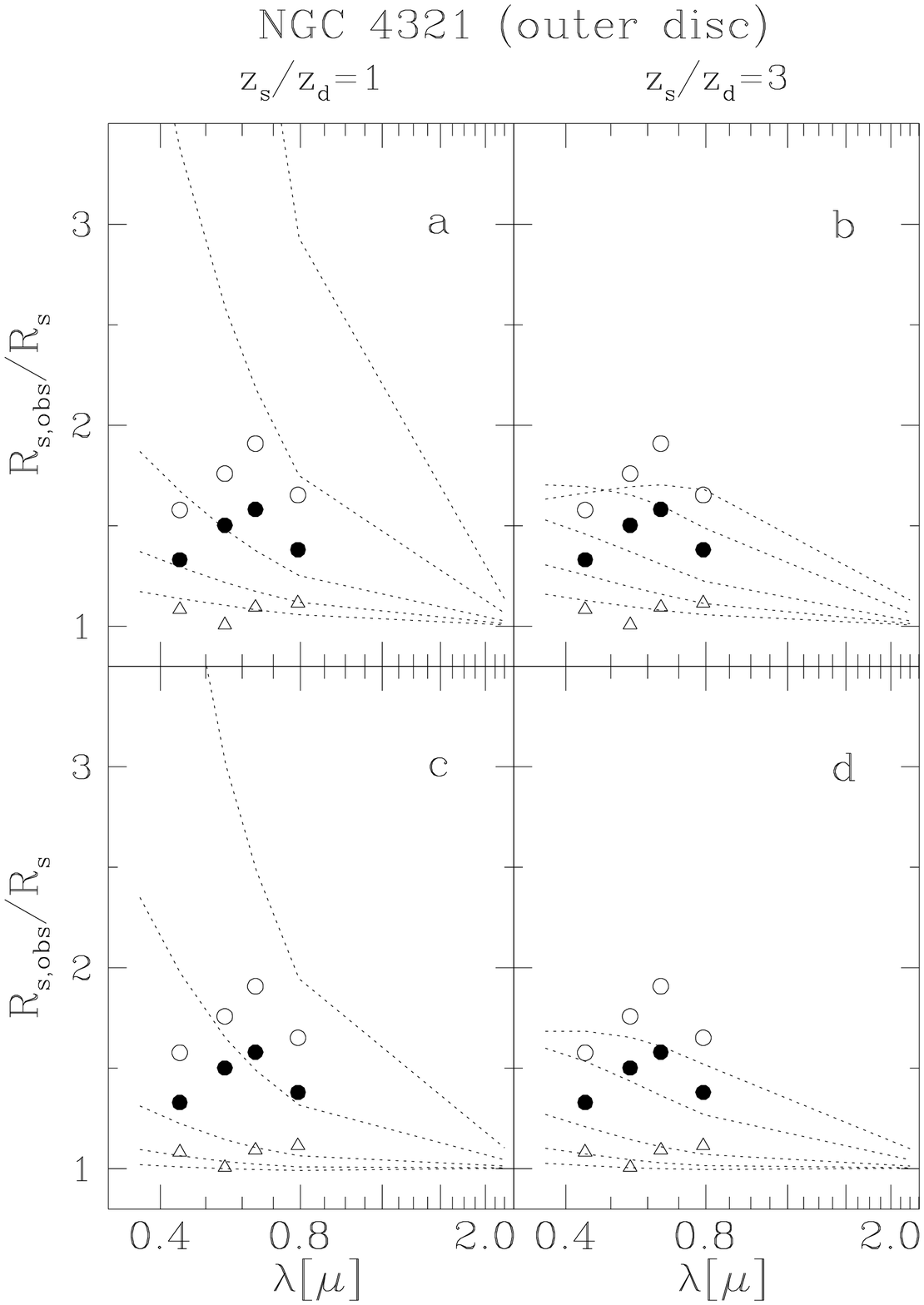}}
\caption{
Scattering models of NGC~4321 for the outer range. Plotted are scale-length
ratios for the interarm region (open triangles), the arms (open circles)
and the whole disk (filled circles), together with model curves for
$\tau_V$ = 1, 2, 4, 10 and 20. The upper models include absorption only,
while the lower models include absorption and scattering.} 
\end{figure}
\begin{figure}
\centerline{\epsfysize=13cm \epsfbox{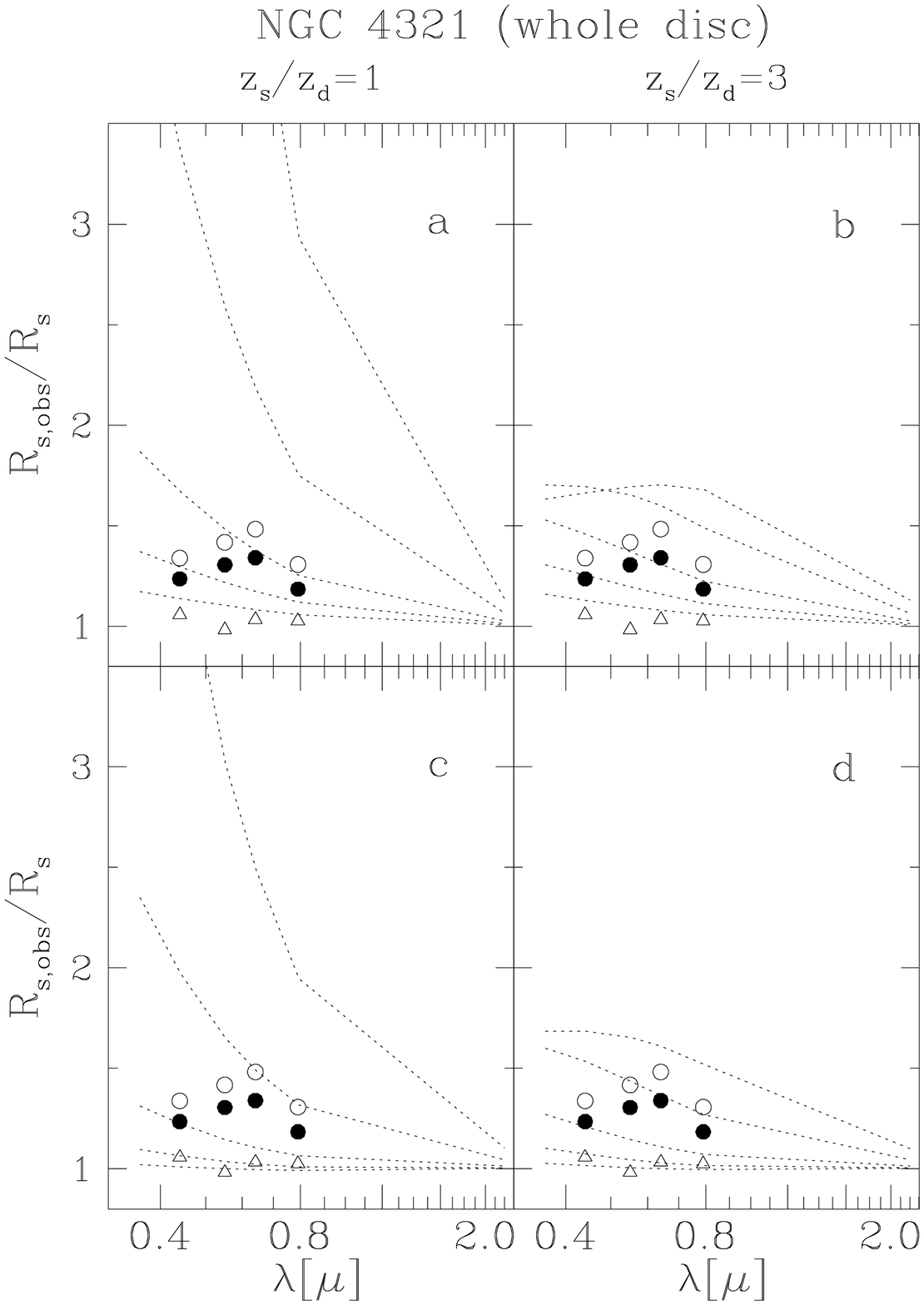}}
\caption{
Scattering models of NGC~4321 for the whole disk. Plotted are scale-length
ratios for the interarm region (open triangles), the arms (open circles)
and the whole disk (filled circles), together with model curves for
$\tau_V$ = 1, 2, 4, 10 and 20. The upper models include absorption only,
while the lower models include absorption and scattering.} 
\end{figure}

\section{Discussion and Conclusions}

We have obtained azimuthally averaged radial intensity profiles in the
$B$ through $I$ photometric bands, of three nearby spiral
galaxies: M~51, NGC~3631, and NGC~4321; for the first of these we also
obtained a profile in the $K$ band. 
As the arms and the
interarm disk zones are manifestly different photometrically, we went
on to derive separate profiles for the two types of zone in each
galaxy, and measured the radial intensity scale-lengths, separately,
for the arms, and the interarm zones, as well as for the galaxy
disks 
as a whole.  The results for M~51 and NGC~3631 show scale-lengths which
decline systematically, but not steeply, with increasing wavelength,
and arm scale-lengths which are longer than the interarm scale-lengths at
short wavelengths, but converge to the interarm values at longer
wavelength.  For NGC~4321 we find  an inner zone which behaves in the 
same way as the disks of the other two objects, and an outer zone
which contains less dust. 
However, our observations go out
further for NGC~4321, and we might well expect similar results 
in the outer parts of the other two objects.
We have interpreted these results using radiative
transfer models which take into account scattering as well as
absorption by dust.  
We characterise the dust distribution by an
integrated optical depth, $\tau_V$ along the central axis of the disk,
as well as by a vertical scale-height, and an intrinsic radial
scale-length.  Since $\tau_V$ 
is a quantity extrapolated inwards to the axis, it
represents an upper limit to the values found everywhere in the disk.
For M~51 the best model fits to the observed scale-lengths yield values
for $\tau_V$ of close to 10 for the arms, close to 1 in the interarm
disk, and a mean of 3 over the disk as a whole.  The corresponding
values for NGC~3631 are 10, 4, and 6 respectively.  

As well as being
upper limits these values yield a misleadingly high impression of the
effect of the dust extinction because we tend to think of an optical
depth in terms of the most familiar scenario in our own Galaxy where
the source lies entirely behind the absorbing dust cloud.  The
$\tau_V$ in the models used here is an integrated value through the
center of the (extrapolated) disk, but the stars and dust are distributed
throughout the disk, so the net absorption will be much lower than if
all the dust were between us and the stellar component.  As an example
for $\tau_V$ = 5 (the on-axis value), and a stellar to dust scale-height ratio of 1, the
dust extinction for the absorption + scattering case
for the face-on system along its axis is 60\% of the total
light. For a ``screen" model (all the dust in front of the stars) the
equivalent would be 99\%.  The corresponding figures for a stellar to
dust scale-height ratio of 3 are 50\% and (of course) 99\%
respectively.  We may also include, for comparison, models with pure
absorption and no scattering.  For $\tau_V$ = 5 and the scale-height
ratio 1, extinction would take out 80\% of the light, and for a scale-height
ratio 3 this would be 67\%.  These figures are larger than
those for the models including scattering, but they should not be
taken as representative of real dust. It is also instructive to make a
further calculation: of the fraction of light removed by dust averaged
over the disk. To be specific, we have integrated the transfer
equations out to 3 scale-lengths from the axis.  Setting $\tau_V$ = 5
and with the stellar to dust scale-height ratio unity, we find that
only 15\% of the total flux has been removed by dust, this figure is
virtually unchanged if we change the scale-height ratio to 3.  It is
notable that in a screen scenario the figure would be 70\%.  From the
numbers presented in the previous paragraph we can conclude that the
light extinguished in M~51 and in NGC~3631 by the presence of dust in
their disks will be of order 15\% of the total starlight
emitted. 
The qualitative conclusion is
that, for these two galaxies, it is most unlikely that a major fraction
of their light is being extinguished by interstellar dust.  

A further
consideration to be taken into account is the non-uniformity of the
geometrical distribution of the dust, notably its compression into
lanes.  It is clear from any image of typical disk galaxies, such as
those in Plate 1, that the dust is in fact concentrated in lanes,
especially prominent along the edges of the spiral arms, where the gas
is being compressed by the passage of a density wave. We have run
models to test for the effects of a dust lane on the scale-length of
(say) an arm. To first order we find that a dust lane which occupies a
fraction x of an arm, with the rest of the arm dust-free, yields the
same effect on the scale-length as if the whole arm were characterized
by an optical depth $\tau_V$, provided that the on-axis extrapolated
optical depth of the dust-lane is $\tau_V$/x.  A modeled lane
occupying only 30\% of the surface of an arm will give essentially the
same averaged photometric arm profile as if the arm had dust uniformly
distributed across its surface, but the dust lane requires some 3
times the on-axis optical depth as the uniformly dusty arm. These
considerations will be dealt with in more depth in a paper devoted
purely to the modelling technique.  However we can infer here that the
organization of the dust into lanes, and the details of such
non-uniformities, would not alter the qualitative conclusion that
there is only moderate to low net visible extinction across the disks
of M~51 and NGC~3631.  
One should note that a
somewhat similar technique to measure the amount of extinction was
used by D.~Elmegreen (1980). She used UBVRI colors of dusty regions in the 
disk of a few face-on spiral galaxies, and compared them with nearby
regions without much dust. Applying a radiative transfer code that
includes absorption and scattering she found values for the visual
absorption of 0.3 -- 1.2 mag in the interarm region, and 1.5 -- 4 mag
in the arms. These values are consistent with what we find here. The 
difference here is that we also require that the radial density
distribution of the dust is exponential, which means that the models
are somewhat better constrained.

The results for NGC~4321 are more complicated because a more 
extended part of the disk has been observed. 
In the inner disk, 
we find similar results to those of the other two galaxies. The
measured scale-lengths, both in arms and interarm zone,
decrease with increasing wavelength, but the systematics can be 
well modeled using dust. The outer disk is best described as a 
system low in dust, and in both inner and outer disk the arms 
show longer intrinsic scale-lengths than the interarm zone, as 
might be predicted from the active star formation  evident in the arms.
It would also be possible to take into account a fractional dust-lane 
cover in the inner zone of NGC~4321 in just the same way as
that described above for M~51 and NGC~3631, and with a similar conclusion.

The overall results of the observational study presented here are
thus:

\begin{itemize}
\item
It is feasible, 
valuable, and physically instructive, to separate the arm regime
from that of the interarm disk when measuring the radial luminosity
profile of a disk galaxy, and to extract separately the respective
scale-lengths, whose wavelength dependence is to be analyzed. In
general the arm scale-lengths are longer than those of the interarm
zones.

\item  In two of the three slightly inclined galaxies observed
we find a simple coherent picture of the wavelength dependence of the
scale-lengths in terms of dust extinction.  The dust optical depth is
measurably greater in the arms than in the interarm disk, a not
unexpected result.  Our model fits imply that the dust removes less
than 20\% of the emitted starlight in the $V$ band from those portions
of the disks within two intrinsic scale-lengths from the axis (at
greater radii this figure would be even lower).
In the third galaxy, where we have been studying 
a greater radial range, the inner
disk shows similar dust properties to the disks of the other two galaxies,
while in the outer disk the most reasonable interpretation of the results
is that there is not enough dust to have a significant effect 
on the scale-lengths.

\item 
The results would remian essentially the same if we assumed that the 
dust was compressed into lanes  occupying a fraction x of the surface area
of the zone under consideration, with axial optical depth greater
by a factor 1/x than for the uniform case.

\item Better fits of the scale-length vs. wavelength observations
are obtained for models with stellar to dust scale-height ratio of 1
rather than 3, notably for the arms. This tentative result would 
be most interesting to confirm with a broader data base.
\end{itemize}

The present study should be seen as a trial in two senses.  Firstly we
have measured only three objects: one Sbc galaxy, and two Sc
galaxies.  There are observational reasons to believe that there
may be less dust 
in later-type galaxies, and any work aimed at drawing
statistical conclusions should have a more representative base, simply
in terms of numbers of objects, but also in terms of the spread of
morphological types. Secondly we have not explicitly taken into account
radial variation in the intrinsic colors of the stellar populations.
The fact that the observed radial color variations are small in these
objects suggests strongly that such stellar effects are not important
here (ruling out the fortuitous cancellation between dust reddening
and stellar population blueing as a function of radius), but any
tendency of the stars in the outer parts of disks to be bluer would
tend to make us under-estimate the dust content.  Here again, with a
wider data set, where comparisons between galaxy types can be made, we
may obtain a clearer idea of how well such population effects can be
quantified.  
In spite of these limitations, we feel that we have shown, in this article,
how the scale-length test may be employed with considerable 
value as a dust diagnostic in disk galaxies.

\acknowledgments
This paper is based on observations obtained in service time
at the Isaac Newton
Telescope and the Jacobus Kapteyn Telescope at La Palma,
operated by the RGO at the Observatorio del Roque de los Muchachos
of the Instituto de Astrof\'\i sica de Canarias. 
We thank H.-W. Rix for making available to us the images of M~51.
This work was partially supported by grants
PB91-0510 and PB94-1107 of the Spanish DGICYT. The work of CRLM is
supported by a grant of the EC Human Capital and Mobility Program.

\newpage

\end{document}